# Performance Evaluation for Privacy-preserving Control of Domestic IoT Devices

**Sameh Zakhary, Thomas Lodge, and Derek McAuley**

Horizon Centre, University of Nottingham, Nottingham Geospatial Building, Triumph Rd, Nottingham NG7 2TU, United Kingdom

Our homes today are equipped with more connected and pervasive device more than ever before. This movement has been accelerated and widespread further as more people continues to work from home a shift that is partially remaining following the COVID-19 pandemic. The Internet of Things (IoTs) devices are scattered all around our environments which are equipped with many types of sensors and can collect, transfer and operate without users' intervention. There are billions of these devices and most of them are either connected to other devices in the home or to external services over the Internet. Most of the existing models for deploying these IoT ecosystem involves the vendor being in the loop of the command and control of these devices hence users' privacy and security is one of the main requirements. Despite these concerns, users are often faced with a choice between limiting the device functionality or enabling internet access to the IoT devices by signing up to the vendor centralized model to be able to access their device from outside their home. In this paper, we argue that although IoT is promising a revolutionary way of offering services to users, most of these devices shouldn't be allowed to have Internet access due to the increased risks to privacy and security. We present an alternative home networking design model which limits the exposure of IoT devices, and enable seamless access to their functionality from outside the home using WireGuard (WG) -a state-of-the-art Virtual Private Network (VPN) protocol-. We built a test-bed using of-the-shelf IoT device for testing our proposed network design under various conditions including access from Home, 4G, Office and Public Wifi networks. Further, we show that our VPN-based remote access to the IoT device offers a better performance in terms of end-to-end delay in all scenarios when using Hypertext Transport Protocol (HTTP). and comparable performance when using double encryption Hypertext Transport Protocol Secure (HTTPS) over the VPN.

**Index Terms**
Home networking, Internet of Thing, Tangable security, Public/private key security , VPN

## I.    INTRODUCTION

The number of connected devices including IoTs on the Internet is growing fast. According to recent Gartner research, the estimated number of IoT devices is 5.8 billion in 2020 (Gartner, 2019). The countries that are leading the way to IoT deployment include North America, Western Europe and China (Kandaswamy and Furlonger, 2018). By 2024, the number of Machine-2-Machine (M2M) connections between these devices are expected to reach 27 billion in 2024 (Kandaswamy and Furlonger, 2018). This growth in M2M connectivity is expected to result from wide range of application areas such as smart cities, smart infrastructure, smart energy among many others (Hassija et al., 2019). This wide spread of IoTs has sparked significant research interest to understand various implications (Airehrour et al., 2016; Neshenko et al., 2019; Hassija et al., 2019). IoTs enable the integration between many objects in our daily life (Aazam et al., 2016; Alaba et al., 2017) such as sensors, objects, wearable devices and other types of machines. IoT devices are capable of communicating directly with one another and sharing data without direct human intervention (Crabtree et al., 2018). These "things" could be any traditional objects such as home appliance (e.g. microwave, fridge) or tiny sensor (e.g. humidity or health sensors). The devices are capable of constant collections of various sensitive and personal data about many aspects of our lives due to its pervasive deployment (Ren et al., 2019).

The number of recent security and privacy issues found with many of these "things" have raised the concerns of end-users as well as various other parties (e.g. campaign groups and governmental departments). In addition to this, the miss-use of these technologies by malicious actors both from within and from outside the domestic settings has lead to various high profile incidents (such as cases of domestic abuse Lopez-Neira et al. (2019); Janes et al. (2020)). The issue is further exacerbated when this "thing" is accessible over the Internet (such as security web-cameras), where an undiscovered vulnerabilities could by exploited by malicious adversaries without the end-user knowledge or before the manufacturer could produce a patch to mitigate exploitation.

Further more, users rely on the vendors' supplied application to control their devices, which usually relay on a Trusted Third Party (TTP) to act as a mediator between the mobile device and the IoT. For these applications to work from outside the home, users have to accept -usually long and difficult to understand- list of terms and conditions in order to enable this feature. Not only this raises a wide range of security and privacy problems (e.g. eavesdropping, data shipped to outside jurisdictions, enabling device manufacturers to collect detailed usage information from the IoT devices), this also introduces a single point of failure (i.e. the TTP). In most cases, the users are not presented with an alternative options for using the advanced application features if they do not wish to enable Internet access to their IoT devices (e.g. users are not able to access/change setting or control their device when outside their home).

In this paper, we argue that most of the IoT devices in-fact do not need to be accessible over the Internet (neither outgoing nor incoming traffic direction). This is a generally standard enterprise network access control policy where devices are categorized and assigned the right policy to allow for proper functioning. We aim to form a weak form of isolation, which is similar to the stronger and more established defence through an"air gap" network security measure where different computer networks (e.g. secure and unsecure) are kept physically isolated from one-another hence not allow threats to propagate between them. This is a difficult challenge —even for enterprise networks with fully dedicated Information Technology (IT) staff— partly due to the massive shift to home working, lack of clear security boundaries between trusted and untrusted networks and the dynamic nature of IoT devices accessing the different networks leading to wide adoption of zero trust models Haddon and Bennett (2021).

Our research is investigating alternative means of deploying and controlling IoTs device in the domestic settings while still allowing users the convenience of controlling their devices from outside the home using VPN technologies. Hence the main contributions of this paper in addressing these key challenges in the domestic network settings are as follows: 1) Propose a micro-segmented network design with the objective of isolating IoT devices and reduce risks of further exposure if one gets compromised, 2) Understand the applicability of existing enterprise VPN solutions and networking tools to achieve better security when deploying IoT, 3) build a testbed and develop a set of scenarios to evaluate the end users' experience impact (i.e. through measuring end-2-end delay)

This paper provides an overview of the literature relating to securing IoT with an emphasis on usability from a user perspective as well as approaches to securing access to these devices over the Internet. Although IoT deployment occurs in various settings, i.e. industrial IoT deployment, we mainly focus in this paper on private residential home deployment (i.e. consumer IoTs). We assume that in such settings, users are mostly not experts in security IoT or the underlying networking principles.

This paper is organized as follows: section II discusses various protocols and networking security tools (e.g. firewall and VPN). In section III, we present a number of existing including enterprise-grade solutions that could be adopted to secure remote access to IoT devices in domestic settings. In section V, we present an overview of the testbed which we used to evaluate the performance of our proposed VPNbased access to the IoT as part of the network hierarchy and various control scenarios. In section VI, we present the evaluation results of each of the scenarios presented in the previous section. We finalizing our paper by presenting a conclusion and future work.

## II.     RELATED WORK: IOT NETWORKING

A number of messaging protocols have been developed to meet the various requirements of IoT deployments (Naik, 2017). There are mainly four messaging protocols widely accepted for IoT systems: Message Queuing Telemetry Transport Protocol (MQTT), Constrained Application Protocol (CoAP), Advanced Message Queuing Protocol (AMQP) and HTTP. The previous protocols sits above the transport layer (TCP/UDP) protocols. Unlike the Internet which has a single standardized messaging protocol HTTP, IoT has different messaging protocols as currently no single protocol is able to meet all the applications requirements Naik and Jenkins (2016).

A.  Network traffic control mechanisms

There exists many networking security tools that can be used to control the follow of traffic in a network (Mortier et al., 2012; Neshenko et al., 2019). Virtual Private Network, firewalls, Wireless Local Area Network (WLAN) security standards (Wireless Protected Access 2 (WPA2) personal & enterprise) Mortier et al. (2012) show that domestic users seek to better understand and control the different flows within their home network. With the advancement and wide spread of IoTs, home networks are faced with a much greater challenge to secure these many of these devices (Neshenko et al., 2019). Researchers have been assessing the security issues associated with some of these IoT devices, and indeed many have been found to have major security flows (e.g. self-signed Secure Socket Layer (SSL) certificates, default credentials or out-of-date software) (Costin et al., 2014). Controlling access of outsiders (i.e. through the Internet) is one technique employed to prevent remote exploitation of some of these vulnerabilities and reducing the attack surface of some of a IoT devices.

Home users are usually offered to configure a simple front end firewall interface on a local webbased interface running on the home router or a cloud-based portal. In the case where the home router is running a linux-based operating system, this mostly translates into an iptables (netfliter) rules to control inbound/outbound traffic flows. Linux iptables/netfilters is one of the most commonly and well established mechanisms for packet filtering for decades. More recently, extended Berkeley Packet Filter (eBPF) had gained more attention as an alternative for offering packet filtering that meets future server deployment and virtualization scenarios that require additional scalability as the iptables rules significantly grow (Bertrone et al., 2018).

Guest WLAN[1] is a feature that exists in commercial as well as open-source router firmware such as OpenWRT Operating Systems (OSs). Guest WLAN uses wireless protocols standards -IEEE 802.11 (WiFi)- to allows connected devices to have access to the network but in a more restricted way. The objective is to create a Guest network that is easy to connect to by devices that are largely isolated from the main network. Most guest networks will provide access to the Internet, while restricting access to the Local Area Network (LAN) in order to minimizes the impact in-case one of the connected devices is compromised. Further more, some guest networks could be provided so devices on this network will be prohibited from routing any traffic whether on the WLAN nor the LAN, further restricting how a malicious device can attack other devices on the guest network. Guest network could be subject to additional monitoring and Quality of Service (QoS) restriction to avoid miss-use (e.g. a malicious device consuming significant bandwidth and causing starvation to other devices on the LAN).

Additionally, domestic routers usually do not offer ways to monitor bandwidth usage by each devices nor the ability to control the usage. This could become an issue when IoT start contending for bandwidth or trigger high bandwidth operations (e.g. OS update, sync operations) that impacts other interactive or real-time services (e.g. live streaming or gaming). For these situations, traffic shaping (also referred to as packet shaping) offers a useful means to control and limit the network bandwidth usage of devices or traffic type. Traffic shaping enables delaying or restricting the rate of a specific type of network packets according to a defined priority which translate to limiting the bandwidth consumed by the associated traffic. These features are mainly targeted towards enterprise grade solutions and high-end routers where it ensures a better QoS for business-related traffic and interactive applications.

---

[1] https://openwrt.org/docs/guide-user/network/wifi/guestwifi/guest-wlan

B. Solution for securing remote access

Completely blocking Internet access to IoT devices or limiting its interactions to only in-door environment might be possible for some category of device, but other IoT devices such as CCTV monitoring systems might need to be monitored while occupants are away from their home. There exists multiple solution offering VPN-related (mostly used in enterprise network to enable remote or tele-working) that could enable secure access to local network resources. VPN approaches could be achieved using various existing security protocols and tools such as IP Security (IPsec), OpenSSH, OpenVPN and WG.

IPsec follows a layering architecture that enables each of the layers to deal with a dedicated set of protocol functions (i.e. key exchange, data transport, encryption, interface, etc.) It relies on using Linux transform ("xfrm") layer which allows users to select cipher algorithm and key in addition to other protocol parameters. The key-exchange stage (usually using IKEv2 (Kaufman et al., 2010)) of the protocol updates various data structures which is used by the following data transport stages. IPsec protocol is complex to setup correctly and requires additional firewall semantic and configuration for it to function behind a firewall.

To avoid some of these issues, other approaches have been developed for VPN. For example, OpenVPN is TUN/TAP based solution running in the user space (e.g. not inside the Linux kernel). It offers a VPN solution that relies on Transport Layer Security (TLS) and is available on many Linux OS distributions. It requires a running daemon responsible for managing the logical network interface (tun0). OpenVPN is quite complex due to supporting many of the TLS features and functionality. TLS is difficult to manage due to its huge state machine.

Donenfeld (2017) developed WG, a secure OSI layer 3 VPN protocol. It is a relatively new VPN protocol that has recently been merged into the Linux kernel V5.6 (Torvalds, 2020). WG offers faster throughput and modern cryptographic support as well as much lower overhead which offers better battery life. As WG is mostly stateless, it is able to support better roaming compared to state-full solutions. As it has been merged into the Linux kernel, it offers a much lower overhead and much higher bandwidth compared to user-space VPN options. Key distribution is not covered by WG protocol, and WG implementation in Linux is mostly agnostic to how users would exchange cryptographic material. This is a similar approach to OpenSSH, where users can exchange their public keys Out-of-Band (OOB) to setup secure connectivity. It worth noting that these keying material is used to drive other ephemeral keys during the actual communication over the VPN tunnel. We will look at WG in further details in section II-C.

Some of the VPN solutions we discussed above have been analysed and compared in the literature. Pudelko et al. (2020) have compared the network performance of OpenVPN, Linux IPsec and WG. Author show that WG has demonstrated the best performance in terms of high-throughput under their proposed performance measurement scenario (site-to-site setups) using pipelining. Since key exchange required for setting up the public keys for the peers is not covered by the protocol offers, we are performing additional research to exchange/drive these keys based on tangible interactions in the domestic settings in this project.

C. Wireguard: Kernel-space state-of-the art VPN

Donenfeld (2017) has developed WireGuard, a secure VPN protocol that incorporated modern cryptographic algorithm. Similar to other VPNs protocols, WG offer a secure communication over insecure channels to authenticated parties (called peers). WG doesn't have the concept of a client and server, hence the use of the name peers, but instead have Initiator and Responder to refer to the peer that starts the communication, and the peer that accept the communication respectively. These roles could change at any stage using WG, so a given peer could be the Initiator at one time, but later becomes the Responder when it gets contacted at a later stage and vice versa.

WG has gained considerable attention in research (Lipp et al., 2019; Pudelko et al., 2020) as well as the wider networking community with its merge into the Linux kernel (Torvalds, 2020). It offers a number of advantages over other existing and well-established VPN solutions, mainly its high performance and simple code base. The fact that WG is developed and integrated directly into the kernel space (not clearly separating the functions performed to adhere to the Open Systems Interconnection (OSI) layers (Standardization, 1996)) enabled many performance advantages including reduced overhead in coping packets across different processes (residing in various kernel and user spaces).

WG provides a secure virtual interface at the OSI layer 3 which relies on Cryptokeys routing. This term is used by WG to refer to the secure and cryptographic-backed mechanism by which the interface will authenticate incoming packets to peers' keys and encrypt outgoing packets with the key associated with the corresponding peers public key. The two peers form a tunnel using User Datagram Protocol (UDP) for transporting packets for all network flows arriving from higher layers addressed to the give WG interface. The packets arriving from upper layers to an address managed by this interface (at L3) are securely encrypted and encapsulated over in UDP packets to the latest known external peer's Internet Protocol (IP) address. Authenticated Encryption with Additional Data (AEAD) schema is used to protect data sent over the tunnel by providing confidentiality, integrity, and authenticity. In brief, the IP packets arriving at the interface are encrypted/signed using ephemeral keys and encapsulated into a newly created UDP packet and WG port of the corresponding peer's WG interface using the last known IP of this peer-. These processes are done behind the scene with the user not involved in tracking peers' IP, session initialization including producing derived keys nor triggering re-keying (i.e. future Handshake process after various timeouts events).

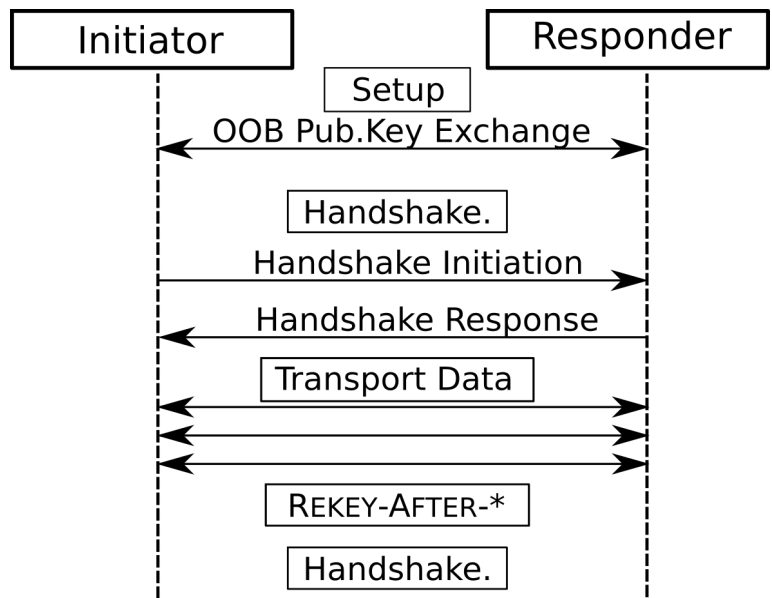

Fig. 1: WireGuard high-level typical sequence of operation between two peers (setup including OOB key-exchange required once for new peer).

Figure 1 shows the high-level sequence of operations during setup and later communication between peers. To setup a WG peer, the two entities need to exchange 32-byte Curve25519 public keys out-of-band. WG doesn't state how this could be achieved, and for the purpose of our paper we assume that keys are exchanged securely through a tangible interaction (as highlighted in section IV). This enables each of the peers to be uniquely identified by its public key, this further protects peers from Denial of Service (DOS) attacks as peers will never respond except if the sender proves that he/she knows their public key. Although public keys are not secret, they are not published and do not form an identity (i.e. one peer can generate a public/private key pair to be used with another particular set of peers on a specific network interface). At the start of every communication, the peer that wishes to send data (called the Initiator) starts the Handshake stage, and the other peer verifies and only respond to legitimate Handshake Initiation packet with a Handshake Response.

Handshake stage is performed using the same UDP protocol port number as the one used for data transport. During the 1.5-Round-Trip Time (1.5-RTT), the Initiator can start sending encrypted messages as soon as it receives the handshake response, while the responder may only send encrypted data once it has received an Acknowledge (ACK) from the Initiator confirming that it has indeed received the response. This is in order to prevent replay-attacks, where an adversary could replay older handshake messages to cause the responder to regenerate its ephemeral key and invalidating the key of the legitimate Initiator. To mitigate against such attack, WG includes 12-byte TAI64N (Bernstein, 1997) that is encrypted and authenticated. The greatest time stamp is stored by the responder for values received from each peer so as to discard any packet that holds values less than or equal to it.

The Noise Protocol Framework (Perrin, 2018) is used during the handshake stage to ensure mutual authentication, key agreement and forward secrecy. This prevents adversaries (unauthenticated or malicious parties) from scanning various networks and hosts for the WG port without knowing the peer long-lived public key. Although WG public keys are not meant to be kept secret (as they get exchanged with other peers), it is important to note that these keys are not expected to become public (i.e. published similar to a Pretty Good Privacy (PGP) key) in order to prevent from DOS attacks.

After a period of communication (timeouts) or a number of messages have been exchanged peers will re-enter into a new handshake stage to change the ephemeral (short-lived) keys currently being used. After this stage, the two peers will resume transporting data using the new keys. After this 1.5-RTT key exchange handshake, the peers can use the agreed pair of symmetric keys (one for sending and one for receiving) to send/receive data over the tunnel.

### III. APPLICABILITY OF EXISTING VPN SOLUTIONS

In the market for enterprise VPN solutions, there many exists vendors. The services are ranging from private to public VPNs. The private VPN could be either on-premises or cloud-based VPN (e.g. VPN as-a-Service) dedicated to a particular client (such as a major bank). Most enterprise-scale network equipment manufactures include VPN, such as Cisco Network Convergence Series Routers. In case of public or shared VPN, the infrastructure is shared to more public VPN solution and is billed as per the usage (e.g. number of connections, clients or bandwidth).

In addition to providing the infrastructure to connect to one of the VPN servers securely, enterprise solutions tends to offer policy and centralized access management to administrators of large organizations (required for compliance and auditing purposes). We have also surveyed some of the key features thought after for enterprise deployment, and we show the features and add important features for home deployment: 1) Ease of deployment into the existing infrastructure, 2) Integration (or compatibility) with IoT, 3) Management and alerting tools, 4) Support during the appliance life, 5) Overall cost (e.g. licensing, support and hardware costs), 6) and Mobility support.

Support for mobile end-devices to access the VPN as the devices get handed-over across networks is referred to as "mobile VPN". For some VPN deployment scenarios, "Mobility" support is not important and might contribute to unnecessary overheads. One example is when a company wants to integrate a remote branches subnets into the larger company network to provide seamless access inside any of the branches. Such scenario requires an enterprise VPN solutions that is focused on wired end-to-end systems with high-speed and reliable infrastructure (referred to as site-to-siteVPN).

On the other hand, "Mobility" support is a corner requirement for most scenarios where an enterprises is enabling remote workers to access local network resources. Also, a usable VPN solution for a domestic network must support mobility as users are mostly roaming on other networks who wish to access their network without degradation of performance or overhead in establishing VPN connections repeatedly.

In addition to the above features, we evaluate the configuration choice supported by each of the possible VPN solutions. The VPN solutions would offer one or more of the configurations listed in table I depending on how the VPN and client are configured. In an enterprise, the choice of which configuration is directly affected by the ownership model of the mobile device which is used to access the VPN. In other words, the level of control and impact on other installed applications on the mobile device could limit which configuration to use for an employee-owned device. For example, an always on VPN" offers the highest security and control for system administrators as it captures all traffic originated from the mobile device and direct it towards the enterprise network. This traffic is subject to the security policy of the enterprise, such as rules of filtering, blocking and monitoring. This is obviously not suitable when the device is owned by the employee who sometimes need to access limited enterprise services (e.g. enterprise email account).

| Configuration | Description |
|---|---|
| **Standard VPN** | Usually requires dialing into a VPN server where all network traffic from the client is routed through the VPN server. This enables monitoring and rule enforcement but causes degradation in performance as the VPN server becomes a single point of failure. |
| **On-demand VPN** | VPN reconnect automatically upon accessing some corporate resource. This enables mobile device to not incur significant delays when accessing other services outside the enterprise. |
| **Per-app VPN** | This VPN configuration allows specific mobile devices' applications to access the corporate VPN server for a specific service (such as emails, file server or sales system). This is similar to the on-demand configuration but focuses on a mobile application instead of a defined network subnet(s). |
| **Always-on VPN** | This configuration enables the device to always be connected the corporate network through a VPN which offers the highest security and control over the devices' network traffic. This enables full monitoring and control over what flows are enabled or which services are accessible to the mobile device. The VPN is started automatically when the device is booted and stays on capturing all traffic (i.e. locking-in the device). |

TABLE I: A possible VPN configuration options.

Alshalan et al. (2016) presented a survey of "mobile VPN" technologies including key software solutions at the date of publication. We revisit the list of enterprise solutions to reflect the recent changes. In this report, we focus on the technologies in the market for enterprise VPN solutions that support Mobility. Based on the aforementioned configuration options, we have evaluated the key players in the enterprise mobile VPN market in terms of the possible configuration(s) and reliance on dedicated hardware/software server in table II.

For consumer cloud-based VPN solutions, there are many vendors offering the user a way to tunnel the network traffic through their servers (for privacy or security when using untrusted networks), for example, NordVPN and ExpressVPN . Looking specifically at available open-source VPNs technology, namely OpenVPN and WG), there exists a paid enterprise or consumer cloud-hosted related service, OpenVPN for business and Tailscale respectively. In table III, we present a short comparison between NordVPN and OpenVPN

There are specific enterprise-level features (e.g. single sign-on, remote access) while other features are useful to all users (e.g. anonymous browsing, peer-to-peer). Single sign-on refers to the ability for users to use they existing enterprise logins to sign-on to the VPN instead of creating separate credentials. Anonymous Browsing is a feature protecting user identity on the web by hiding their real IP (e.g. using onion routing) or tracking technologies such as cookies (e.g. using Tor browser or other plugins). DNS Leak Protection refers to a property whereas the Domain Name System (DNS) related queries and responses are sent through —over a VPN— to DNS servers running in the VPN as opposed to a DNS server provided by the current network. This prevents curious or malicious Internet Service Providers (ISPs) from tracking and recording DNS traffic, hence mitigate against a privacy attack where outsiders are able to know all websites or services the user or devices are accessing. Web Inspection is the ability of the VPN server to inspect web traffic (which usually means performing packet inspection at the application layer—OSI Layer 7—), to detect malware, spyware or other harmful content, and respond to any threat.

| Enterprise VPN | Summary |
|---|---|
| AnyConnect VPN (Cisco Inc., 2020) | <ul><li>TCP-based application access or Datagram Transport Layer Security (DTLS)</li><li>always-on, on-demand or per-app VPN</li></ul> |
| (Citrix Systems, 2020) SSO | <ul><li>layer 3 SSL connectivity</li><li>requires Citrix Gateway</li><li>always-on, on-demand or per-app configuration</li></ul> |
| Mobile Connect (SonicWall, Inc., 2020) | <ul><li>SSL-based VPN</li><li>requires SonicWall Secure Mobile Access (SMA) or Next-Generation Firewall</li><li>on-demand or per-app conf.</li></ul> |
| Connect Secure (Pulse Secure, 2020) | <ul><li>SSL-based VPN solution</li><li>requires a VPN gateway</li><li>always-on or per-app conf.</li></ul> |
| Workspace ONE (VMWare Inc., 2020) | <ul><li>requires Workspace ONE Assist Server</li><li>on-demand or per-app conf.</li></ul> |
| Synopsis (Radio IP Software, 2020) | <ul><li>requires a gateway</li><li>standard conf.</li></ul> |
| (NetMotion, 2020) VPN | <ul><li>requires NetMotion gateway</li><li>always-on or on-demand conf.</li></ul> |
| WireGuard (Donenfeld, 2017) | <ul><li>no specific server (uses peers)</li><li>on-demand (specific WG peers' IPs/keys) conf.</li></ul> |
| (OpenVPN Inc., 2020) | <ul><li>requires access server for business</li><li>or local deployment for home users</li><li>on-demand or standard conf.</li></ul> |

TABLE II: Key enterprise mobile VPN solutions: summary of common deployment option and configurations.

| Feature | NordVPN | OpenVPN |
|---|---|---|
| single sign-on | ✗ | ✓ |
| Anonymous Browsing | ✗ | ✗ |
| DNS Leak Protection | ✓ | ✓ |
| Multi-Protocol | ✓ | ✓ |
| Kill Switch | ✓ | ✓ |
| Peer-to-Peer | ✗ | ✗ |
| Policy Management | ✗ | ✗ |
| Remote Access | ✗ | ✓ |
| Web Inspection | ✗ | ✗ |
| Client/Server Open Source | ✗ | ✓ |
| Deployment | Cloud | Cloud/Private |

TABLE III: Sample comparison of a popular VPN solutions.

Many enterprise solutions exist for the purpose of visualization and obtaining analytics of the network traffic, alerts and other events. We have considered the following examples in this report focusing on open source: Cacti , Nagios , Grafana. These tools enable performing many of the visualization tasks in networks of varied sizes including showing time-series Simple Network Management Protocol (SNMP), examine different services' logs (e.g. for an application or server), and in some cases remotely controlling (e.g. restarting) services. This allows home users or system administrators to configured different monitoring probes or alerts when a specific condition has occurred and possibly automate some actions.

For most enterprise settings, administrator usually need to develop and customize the configuration of various dashboards to monitor different aspects of a large network. On the other hand, for a home environment, users are likely to share specific monitoring and alert configuration for various IoT devices, hence reduce overhead in creating these every time a new device is introduced. User also can access some pre-built (or template) dash-boards, probes or alerts which they can use for their own network.

1) Existing Firewall Solutions for Home Network: As for firewall solutions, there exists many enterprise and consumer grade solutions. Most routers sold to consumers incorporate VPN, firewall as well as other functionality (such as Network Attached Storage (NAS)) —which are typically deployed and managed separately in an enterprise setting— all in a single box (e.g. home router). We have surveyed many existing firewall solutions, and we focused this section on firewalls that can be deployed in domestic settings (i.e. on a router or a general purpose Personal Computer (PC)). Namely, we focus on OPNsense as it is an open-source and the project is under active development.

OPNsense© is forked from pfSense© and has been mostly re-written. pfSense© is a commercial product that has some features released in an open-source version called pfSense© Community Edition (CE). The open-source community edition is released and attracts contribution from a community of interested developers. The enterprise version of pfSense© has license fees. As both of these firewalls have overlapping interested community, the two have been deployed by end-users who wish to customize and add new features. We provide here a high-level comparison of some of the existing firewall solutions as an example. In table IV, we show a brief set of features with a focus on the ability to deploy and customize (add/remove a specific functionality) in domestic setting.

|  | OPNsense | pfSense CE |
| --- | --- | --- |
| License | BSD 2-Clause "Simplified" or "FreeBSD" | Apache License 2.0 |
| Contribution License | BSD 2-Clause "Simplified" | subscribe and electronically sign agreement |
| IPS | Suricata Inline Intrusion Prevention System (IPS) | Snort IPS package |
| Extension | ✓ plugins | ✓ packages |
| Security Update | weekly | patch release |

TABLE IV: Summary of features of OPNsense© and pfSense© firewalls.

## IV. HIGH-LEVEL DESIGN OF DOMESTIC NETWORK TO INCORPORATE IOT

In this section, we discuss a proposed design which deploys WG to facilitate secure tunnelled access into home network. As discussed earlier, initializing and distributing the cryptographic keys is required to enable secure deployment of WG in domestic environment and increase usability.

There are several advantages to using a VPN-based approach to control IoT devices. First and most important to this project is enforcing a secure boundary around the IoT device where external parties have no access to them eliminating the need to inbound traffic from the Internet. For example, when providers expose services directly on the device that could be accessed remotely such as with web enabled CCTV systems. On a more general principal, we see that many IoT devices would not even require an outbound traffic to the Internet and hence a tighter secure boundary can be established. This has the advantage in minimizing exposure of the home network and possibly limit the risk in case any of these devices get compromised.

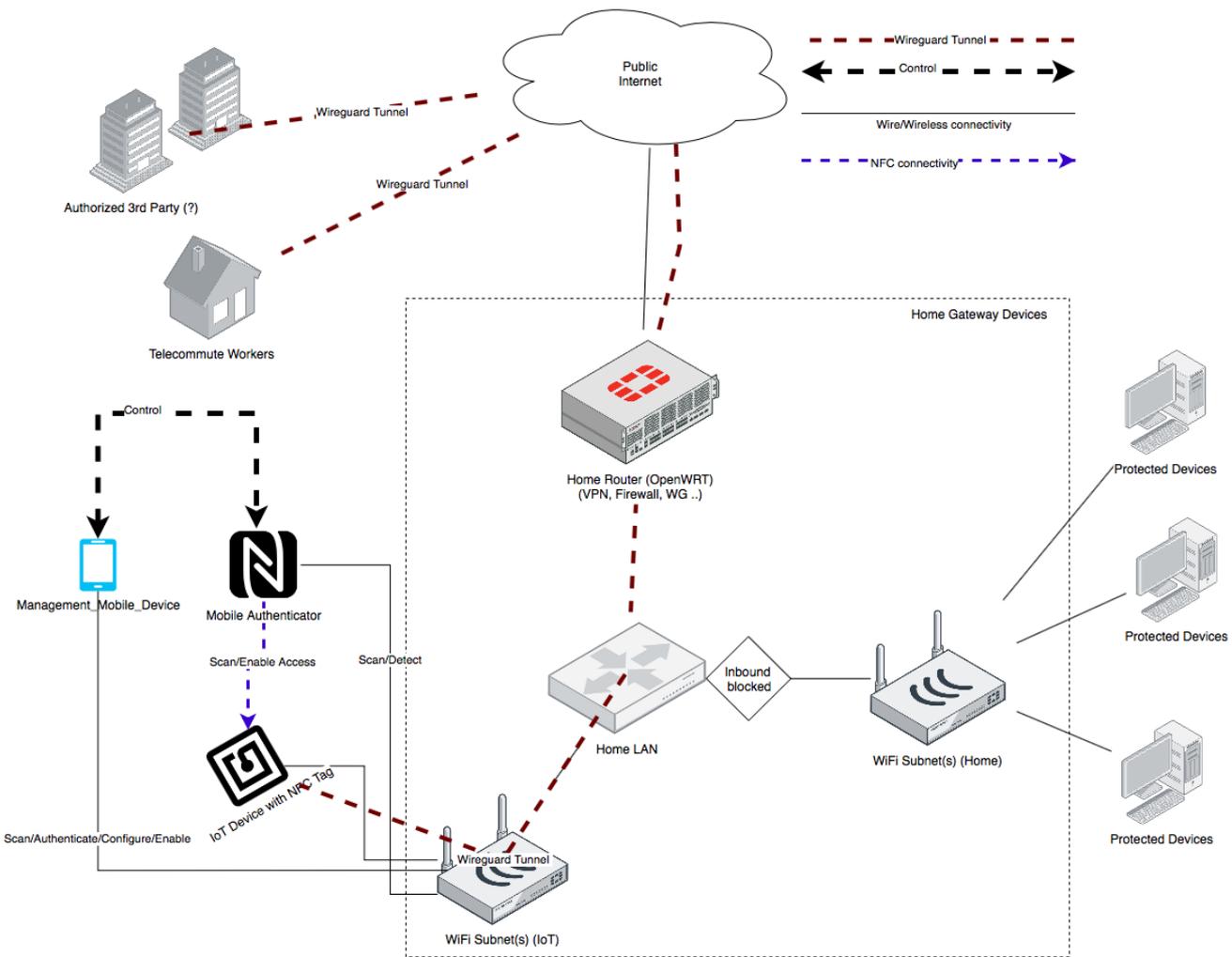

Fig. 2: Overall domestic network diagram using WG to enable secure remote access to IoT devices via tangible interactions.

Figure 2 shows a proposed deployment of WG as a VPN solution. This ensures a secure tunnelling where remote management clients (e.g. the users designated device) is able to monitor and receive alerts related to the various problems occurring in the network. We propose that the secure association between the users' management device and a certain subnet is performed using a token-based NearField Communication (NFC) tangible interaction, which will then enable this VPN access for this device to one or more of the subnets (more discussion on the subnets later). To avoid enabling wide network access, we propose to have the tunnel associated with a particular subnet where the IoT devices will be deployed.

Figure 3 shows our proposed approach to partitioning the domestic home network into multiple subnets. We show -as an example- the corresponding subnetting using IP version 4 (IPv4), CIDR prefix to drive the subnet Mask in figure 4. Classless Inter-Domain (CIDR) enables efficient routing as a system will search locally for the target host if it is within the same subnet, otherwise -if the target is on another network- traffic will be forward to the gateway to perform routing. This would be useful in the scenarios where a domestic users deploy multiple access-points to handle specific subnets where most of the traffic within a subnet is forwarded directly (instead of relying back to the home gateway). As this configuration is relatively complex and requires skill from un-savvy users, it is assumed that home gateway devices will be pre-configured and home users only add any new devices to their respective subnet.

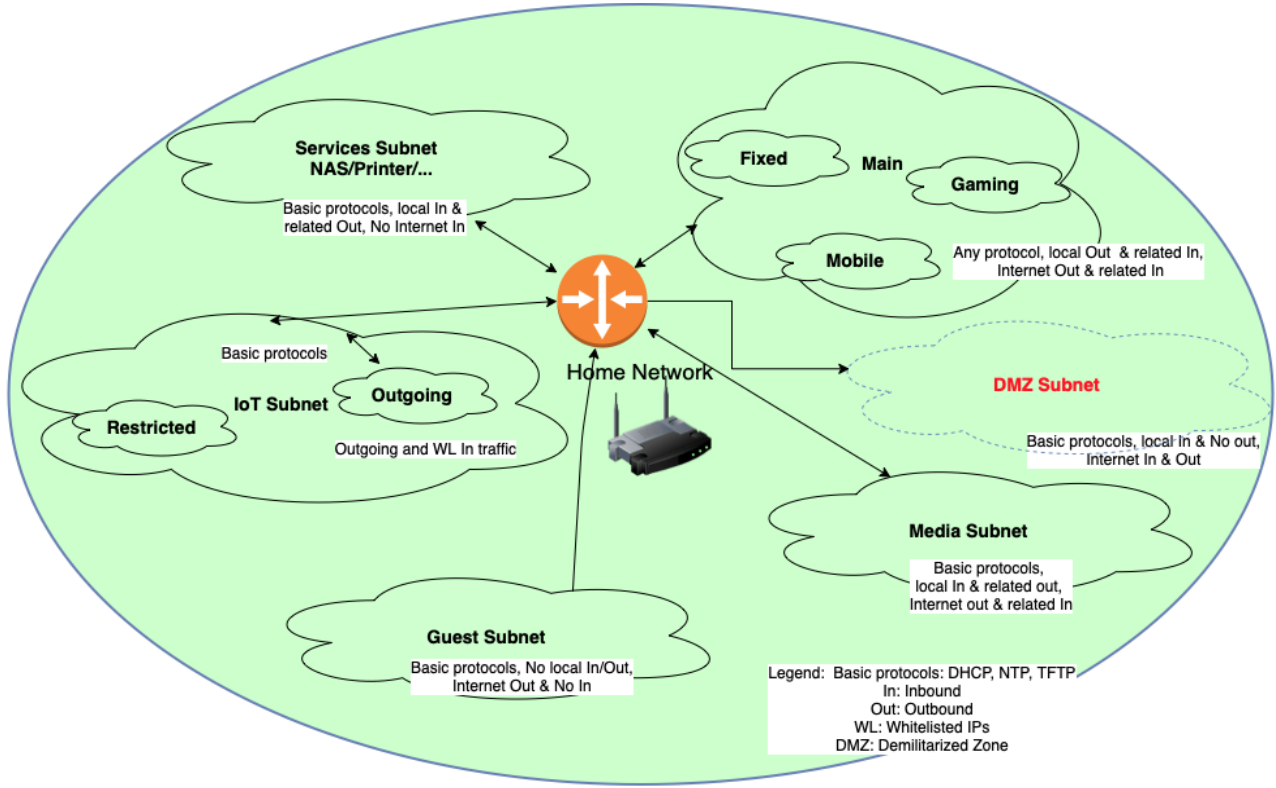

We use IPv4 in figure 4 because it is widely supported by home routers. It is fairly trivial to perform this same exercise for IP version 6 (IPv6)-only network using a single Global ID in the "Unique Unicast" range of fc00::/7 (which is not routable on the global Internet as defined in RFC4193) and assign hierarchically the different subnets in a similar way. Note that we are not using site-local addressing range which were deprecated by RFC3879).

The rationale behind segregating the network into these different subnets is to enable the use of Virtual Local Area Network (VLAN). Most mid-range domestic routers support using VLAN, each of the VLANs would be linked to one of subnets. This ensures that the different subnets are kept separate in management and monitoring, and ensure isolation except where routing and firewall rules permits access.

| Network/Subnets | | | IPv4 | IPv4 (hex) | Prefix | Note |
|---|---|---|---|---|---|---|
| Home | | | 192.168.0.0 | C0.A8.00.00 | /16 | |
| | Main | | 192.168.0.0 | C0.A8.00.00 | /20 | |
| | | Fixed | 192.168.0.0 | C0.A8.00.00 | /24 | e.g. Computers |
| | | Mobile | 192.168.1.0 | C0.A8.01.00 | /24 | e.g. Smart phones/tablets |
| | | Gaming | 192.168.2.0 | C0.A8.02.00 | /24 | e.g. Gaming and other consoles |
| | Services | | 192.168.16.0 | C0.A8.10.00 | /20 | Network attached services (e.g. NAS/Printer/…) |
| | IoT | | 192.168.32.0 | C0.A8.20.00 | /20 | All IoT (no device-2-device comms by default) |
| | | restricted | 192.168.32.0 | C0.A8.20.00 | /24 | No incoming/outgoing (except limited DHCP, NTP, TFTP,…) |
| | | Outgoing | 192.168.33.0 | C0.A8.21.00 | /24 | Enables outgoing connection, and related incoming. |
| | Media | | 192.168.48.0 | C0.A8.30.00 | /20 | E.g. smart TVs, home theaters |
| | Guest | | 192.168.64.0 | C0.A8.40.00 | /20 | Internet only, no access to other parts of the network. (e.g. guests' phone) |

Fig. 4: Home network proposed subnets design to further isolate IoT and other home devices.

## V. EXPERIMENTS DESIGN FOR PERFORMANCE MEASUREMENTS

In this section, we provide an overview of the testbed used to evaluate the performance of our proposed VPN-based access to the IoT as part of the network hierarchy shown in figure 3. The evaluation is to compare the performance between our WG-based control vs. cloud-based approach. We commissioned a simple a IoT system comprising of a Philips Hue Hub and a light bulb. We then issue numerous commands to turn the light on and off, then measure the delay between sending the command and getting a reply from the Hue Hub system or Cloud-based service (in the scenarios where it is used). Figure 5 shows a summary of the various scenarios we intend to evaluate in this paper.

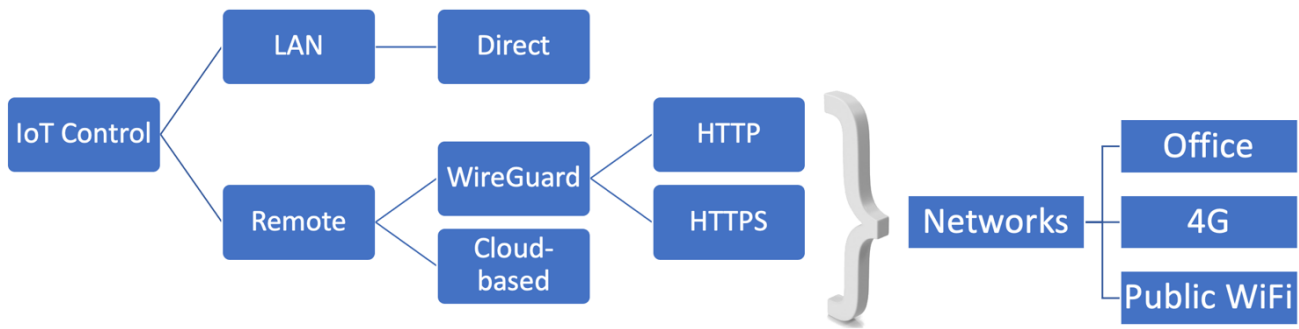

Fig. 5: Experiments design Cloud-vs-WG IoT control scenarios.

The Hub can control several light bulbs via Zigbee protocol. In the case that one of the bulbs is not within the Zigbee range of the hub, then an intermediary bulb act as a relay of the signal to other bulbs (one-hop). We are not concerned with the Zigbee communication delay but aim to measure the overall delay of the system. In order to avoid the multi-hop communication delay, we focus on measuring the delay to one bulb within short proximity of the hub (i.e. no relays are needed).

The communication protocol between the controlling application is over TCP/IP. This control is usually done via the Philips Hue mobile application or any another application that is using the RESTful APIs on the Hue bridge or the Philips Hue Remote APIs. For the latter, end-user must authorize the application on the Philips Hue portal which creates an access token that the application can later use to authenticate when calling the remote APIs.

For the local control scenarios, a secret Application Programming Interface (API) key (sometimes referred to as a user in the Philip Hue APIs documentation) must be created first on the Hue Hub. When the RESTful API to create this key is called, the caller must confirm ownership of the system by means of physical interaction with the system (i.e. pressing a button on the Hub) to authorize the creation. This key must be presented to authorize any future calls to the APIs over the LAN to obtain access to the Philips Hue light system and control it.

We have developed four different scenarios to measure the performance which we are going to describe below.

A. Controlling the light over LAN (local)

In this scenario, the Philips Hue Hub and the controlling PC are located on the same LAN. The actual light is positioned in proximity (less than 1 meter) to the Hue Hub to ensure Zigbee signal is not obstructed. In other words, we are not measuring the delay between the command reaching the Hub.

Controlling commands is issued using (PoSHue PowerShell scripts) to call the bridge RESTful APIs to switch the light on and off. There are two types of delays associated with using an IoT companion application. The first delay is from when the API is invoked until it returns a "success" response to the user, which could be while the command is still being carried (asynchronous execution).The second delay is measured between the command invocation until it has been carried out on the system, i.e. the light has turned on or off since the call was made to the Hub.

The latter form of actual delay measurement (until the light change status and caller is informed) is more relevant to our experiments, but also more complicated to measure. This is because using the companion application, it is not possible to see the data exchanged behind the scene nor record when the light has actually been turned on/off. One way to measure the actual delay was to use a blocking method (synchronous) when calling the RESTful API on the Hub or the cloud services. This way, our controlling device (the client) is blocking until the server has responded to the command either with success or failure.

In order to capture any discrepancy, we also tracked the real status of the light bulb (on/off). We have built a prototype to track the status of the light change using a optical sensor positioned close to the light bulb and isolated from any other outside light source. The sensor is then connected to a monitoring hardware (running over RaspberryPI 4) which is used to detect and record every time light status has changed (on to off, or off to on). In order to match these events to the calls initiated on the control side (the client's system), we had to ensure that systems' local time is synchronized (using Network Time Protocol (NTP)) as the two systems are disjoint and do not have to reside on the same device nor even be located on the same network. As we show later, the controller that issues the command to control the lighting system (hub and lights) at home can be at a completely different physical location.

B.  Controlling the light using Philips Cloud Services (via INTERNET)

In this scenario, we measure the round-trip delay time for the API endpoint to reply to the caller. This is not necessary the time it takes to carry the actual command to turn the light on or off. So, we envisage that there is additional delay of carrying the physical change of the light status which is not captured by the following statistics. The caller is a PC (running PoSHue script) that has Internet access via a Guest WiFi where there is no direct link between it and the Philips Hue Hub. This ensures that all communication to control the Philips Hue light is done through the Philips Hue cloud services.

The Philips Hue cloud services is hosted -at the time of writing- on Amazon Elastic Computing instance (EC2). The first call is made to an IP version 4 address ('52.51.254.112') which is registered under 'Amazon Technologies Inc.' We have attempted to establish the round-trip delay using Internet Control Message Protocol (ICMP) but we received no reply from the server. This is generally due to servers or data-centre administrators blocking this protocol to prevent denial-of-service attacks or mapping the internal network.

It is important to note that the Hypertext Transport Protocol Secure is the only supported protocol to access the Philips Hue REST API cloud services. This ensures that the session and cryptographic material exchanged to authenticate users do not get compromised. Since the communication with Philips Hue cloud services is done over untrusted Internet links, using this application layer security is not optional but important in order to ensure end-2-end private communication. From our delay analysis perspective, this protocol adds further inevitable delays, i.e. the minimum delay achievable takes into consideration additional cryptographic operations carried on both sides of the HTTPS session.

C.  Controlling the light using Wireguard VPN Tunnel – HTTP/HTTPS (over 4G network)

This scenario is similar to scenario V-B above, except that the controlling device is using a separate 4G network that is not connected to the home network. We have constructed this scenario where the WG VPN tunnel is established directly between the controlling device and the home router. The end point of the tunnel (WG peer) is restricted to access only the IoT subnet to minimize the exposed part of the network through VPN. We opted-in to use HTTPS protocol to access the RESTful API on the Philips Hue Hub albite we could have used unsecured application protocol (i.e. HTTP).   *JOURNAL NAME* MARCH 2022 15 The reason for using HTTPS is used because in this scenario we assume that the local subnet where the IoT devices exist is not fully trusted. This prevents one compromised IoT device from easily exploiting other

IoT devices in the same subnet by ensuring that all other devices communications are encrypted too and require prior authorization to intervene with.

D. Controlling the light using Philips Cloud Services (over 4G)

We have constructed this scenario where the controlling device controlled the IoT device via Philips Hue Cloud Services. There was no special configuration done to the home router except to partition the subnets. The IoT devices were isolated in their own subnet with limited access to the Internet. The only protocol to access the Philips Hue cloud service is using the HTTPS protocol where the RESTful API are exposed to control the local Philips Hue Hub.

This scenario represents the only available means for most users to control their IoT devices while outside home. We measure the delay when the user's device is not connected directly to Internet (e.g. via WiFi) but through the mobile network. The users rely on a trusted third-party service offered by the providers' (e.g. Philips in this case). We note that to enable the out-of-home control in the Philips Hue mobile application, the Hue bridge maintains a regular connection with the Philips Hue Cloud Services which is an outbound traffic from the home network. This consumes valuable bandwidth to maintain the connection between Philips Hue hub and Philips Hue cloud service, but we do not measure this overhead between the different scenarios. This is because overhead causing wasted bandwidth is largely dependent on the measurement window and frequency of use which is user-specific.

E. Controlling the light using Wireguard VPN Tunnel – HTTP/HTTPS (over office network)

In this scenario, the Philips Hue Hub is located at the home premises and the controlling device is located at a geographically different location (i.e. office premises in this case). The office is located in at one of the university buildings and is equipped specifically with a high speed Internet connection. This Internet link is not managed through the university's normal firewall so as to avoid restrictions on initiating traffic not going through web-proxy and other policies, while reducing latency. This means that the two devices are not on the same LAN and any communication will have to be done remotely over the Internet. The actual light is at the same location -relative to the hub- as it has been in the previous experiments (i.e. a few meters away). This is to ensure that the Zigbee signal between the Hue Hub and the light bulb is not obstructed.

We have repeated this scenario but this time the controlling device is using HTTPS secure protocol as another layer of encryption above the WG tunnel to communicate with the IoT device at the home network. This attempt to use HTTPS in this scenario following the assumptions mentioned in section V-C.

F. Controlling the light using Philips Cloud Services (over office network)

This scenario is similar to scenario V-E above, except that the controlling device is using the Philips Hue Cloud services to control the light at the home network.

G. Controlling the light using Wireguard VPN Tunnel – HTTP/HTTPS (over Public WiFi)

In this scenario, the Philips Hue Hub is located at the home premises and the controlling device is located at a public space in geographically different location (i.e. public space). The Internet access provided for this WiFi is managed by the City Council and offered at many public offices and leisure centres around Nottingham city. This WiFi although publicly accessible, it still requires authentication through a registration process via mobile phone. The Internet connection speed is not high-speed and firewall rules didn't cause further restrictions to our connectivity. All communication will have to be done remotely over the Internet. The actual light is at the same location -relative to the hub- as it was the previous scenarios. We have performed two sets of experiments using http and https, to compare performance according to the assumptions mentioned in section V-C above.

It is important to note that we attempted to use other open access WiFi (e.g. ones offered by retailers and various restaurants), but we were unable to connect to the Philips Hue cloud service or perform a VPN connection using WG. We suspect that this is mostly due to firewall rules and access policy to prevent miss-use of these WiFi networks as they usually do not require authentication before use.

H.  Controlling the light using Philips Cloud Services (over Public WiFi)

Similar to V-G, but in this scenario we use the public WiFi network to control the light via the Philips Hue cloud services. The communication is allowed through the public WiFi, hence the PoSHue APIs connected directly to the cloud services. Please note that any delay inherent in this public network will be the same delay incurred when comparing the WG-based scenarios, so the comparison is equivalent and not disadvantageous to either case

## VI.  PERFORMANCE EVALUATION

In this section, we show the evaluation results of each of the scenarios presented in section V. The IoT testbed is a Philips Hue light system which was instrumented to perform automated tests and collect timestamps of events. Firstly, we show baseline results measured inside the domestic network where the Hue Hub was positioned for most of the experiments (both controlling the light system directly over LAN and remotely via Philips Hue cloud services.

Evaluation of Domestic Control in Scenarios V-A and V-B

| Delay (ms) | |
|---|---|
| **Mean** | 72.92 |
| **Standard Error** | 0.54 |
| **Median** | 70.67 |
| **Standard Deviation** | 17.22 |
| **Simple Variance** | 296.60 |
| **Kurtosis** | 153.77 |
| **Skewness** | 9.72 |
| **Range** | 373.78 |
| **Minimum** | 28.33 |
| **Maximum** | 402.12 |
| **Confidence Level (95%)** | 1.07 |

TABLE V: Statistics for round-trip delay for locally controlled Philips Hue light system.

Table V shows the descriptive statistics for the round-trip delay measured as the time between calling the Hub RESTful API to switch the light on/off until the return of a successful reply. In the event of an error and the API is failed, the corresponding reading is not reported (e.g. due to system or network failures) as we are only interested in measuring the delay during normal operations. The average delay of the API reply is about 73 ms. It is worth noting that we measured the round-trip delay on the LAN and it is less than 1 ms. As we can see from the 95 % confidence interface -calculated using T-distribution falls within +/- 1.07 ms of the sample mean. We consider the maximum reported delay value in the sample (around 402 ms) an outlier as it is falling far from the sample mean and its associated standard deviation of 17 ms.

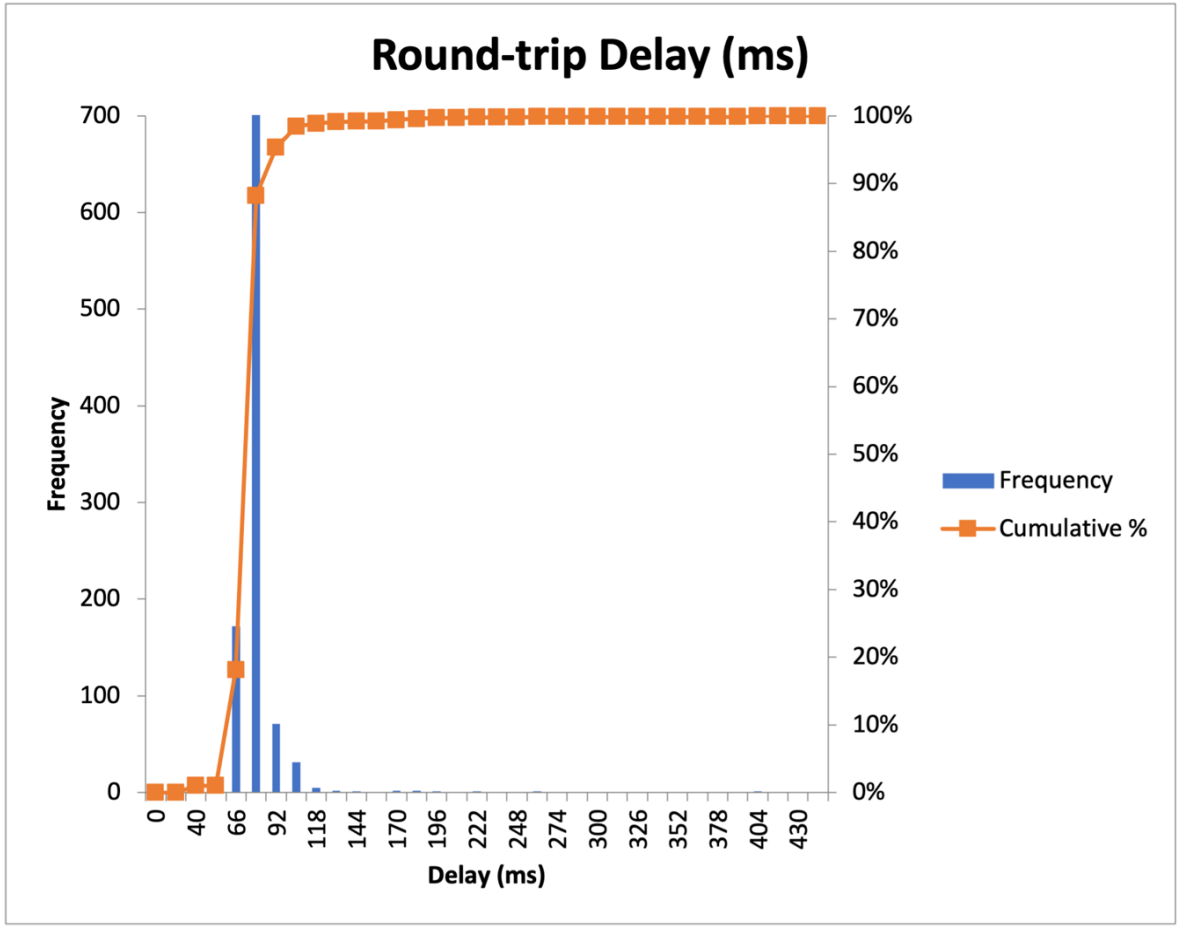

Fig. 6: Baseline: Round-trip delay histogram and cumulative probability for locally controlled light system directly via LAN.

Figure 6 shows the histogram and Cumulative Distribution (CD) of the experiments described in section V-A (i.e. LAN-based). We see that most of the samples falls in the first bin range of below 79ms. The CD reaches 88 % of all values around the mean value with very few values above the mean. The figure shows that 99 % of all recorded delay values are below 130 ms. The bins range (on the x-axis) is chosen depending on the data set range because they are particularly smaller compared to the following scenarios. The CD shows that 97 % of the samples are below 105 ms.

The best performance (minimum delay) recorded in this scenario was 28 ms approximately. This is an important measure as it gives an indication of the best performance possible excluding all the other external factors that might adversely affect the delay (e.g. network congestion, system overload, link failure).

Table VI shows the descriptive statistics of the samples collected. We see that the mean value of the delay is now at 557 ms, which is a much higher compared to the local scenario. This is expected as the command and reply have to travel to the data-centre for processing before a reply is sent back to the caller. As we have little knowledge of how the Philips Hue APIs are implemented, it is difficult to attribute or breakdown what contributes to this delay. We also see a relatively wider value of the 95 % confidence interval here (compared to locally controlled light in table V) around the mean value which suggests that we have a wider variation in delay.

Figure 7 shows the histogram and cumulative distribution of the delay measurement samples when controlling the light via the Philips Hue remote APIs. We can see the shape of the curve being close to a normal distribution with very little skewness to either side. Most of the delay values fall tightly around the mean. The CD shows that 97 % of the samples are below 664 ms, which is much higher delay (worse) than in scenario V-A. The best possible performance (minimum delay) that we recorded in this scenario was 447 ms approximately. This is significantly higher than the minimum recorded delay in scenario V-A of 28 ms.

| Delay (ms) | |
|---|---|
| Mean | 557.05 |
| Standard Error | 2.52 |
| Median | 541.84 |
| Standard Deviation | 79.90 |
| Simple Variance | 6383.28 |
| Kurtosis | 42.27 |
| Skewness | 5.61 |
| Range | 942.02 |
| Minimum | 447.13 |
| Maximum | 1389.15 |
| Confidence Level (95%) | 4.94 |

TABLE VI: Statistics for round-trip delay for remotely controlled Philips Hue light system.

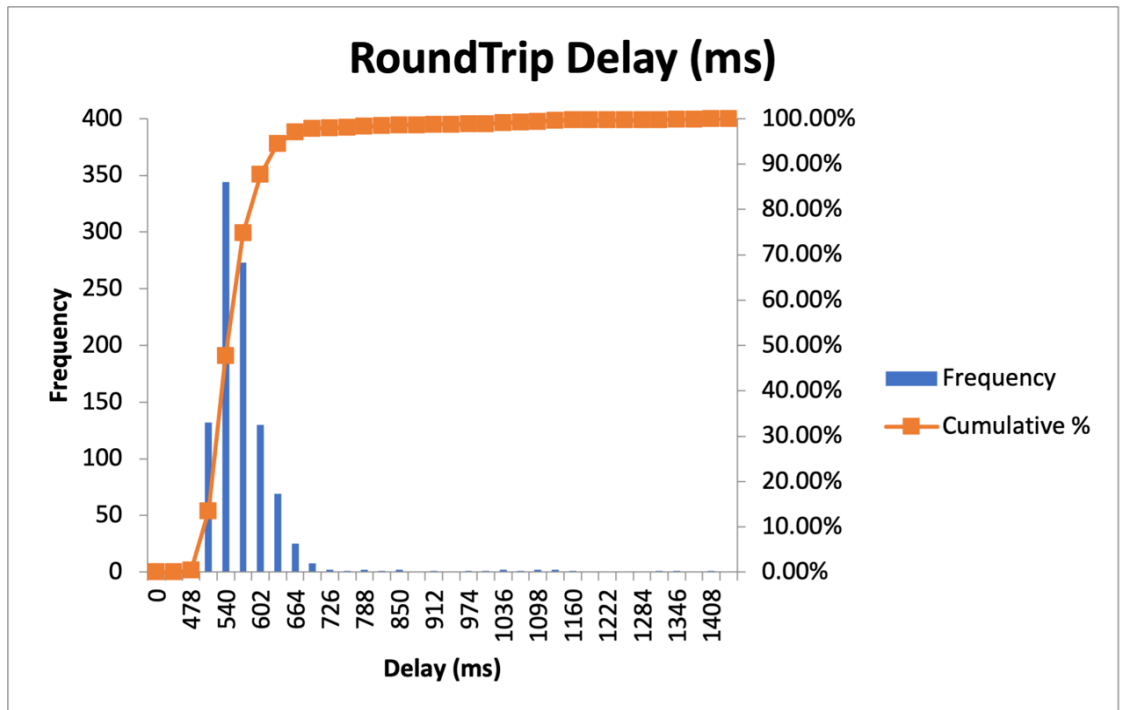

Fig. 7: Baseline: Round-trip delay histogram and cumulative probability for remotely controlled light system using Cloud Services.

Evaluation of Control via 4G Network in Scenarios V-C and V-D

Table VII shows the descriptive statistics of the captured events. For WG-HTTP scenario, we observe a minimum delay of about 309.79 ms which is the lowest achieved for all three. While for WG-HTTPS scenario, we observe more than doubled minimum delay at about 723 ms. The mean follows a very similar trajectories where WG-HTTP achieves the lowest delay albite with a higher standard deviation than WG-HTTPS. The significant overhead in the minimum delay could be attributed to adding two layers (i.e. using both the WG VPN and HTTPS), the first is WG to get traffic to the home router which decapsulate the payload which is another layer of encryption using HTTPS that are then passed to the Hue Hub to decrypted then the command can be executed.

| Delay (ms) | WG-HTTP | WG-HTTPS | Cloud |
|---|---|---|---|
| **Mean** | 369.17 | 948.51 | 938.51 |
| **Standard Error** | 3.40 | 2.80 | 22.78 |
| **Median** | 354.10 | 945.31 | 840.12 |
| **Standard Deviation** | 107.63 | 80.38 | 795.82 |
| **Simple Variance** | 11585.27 | 6460.59 | 633323.88 |
| **Kurtosis** | 215.59 | 62.92 | 82.94 |
| **Skewness** | 12.66 | 5.59 | 9.15 |
| **Range** | 2303.85 | 1226.46 | 7721.95 |
| **Minimum** | 309.79 | 788.98 | 723.11 |
| **Maximum** | 2613.64 | 2015.44 | 8445.06 |
| **Confidence Level (95%)** | 6.67 | 5.49 | 44.70 |

TABLE VII: Statistics for round-trip delay for controlled Philips Hue light system using (WG and Cloud Services) via 4G.

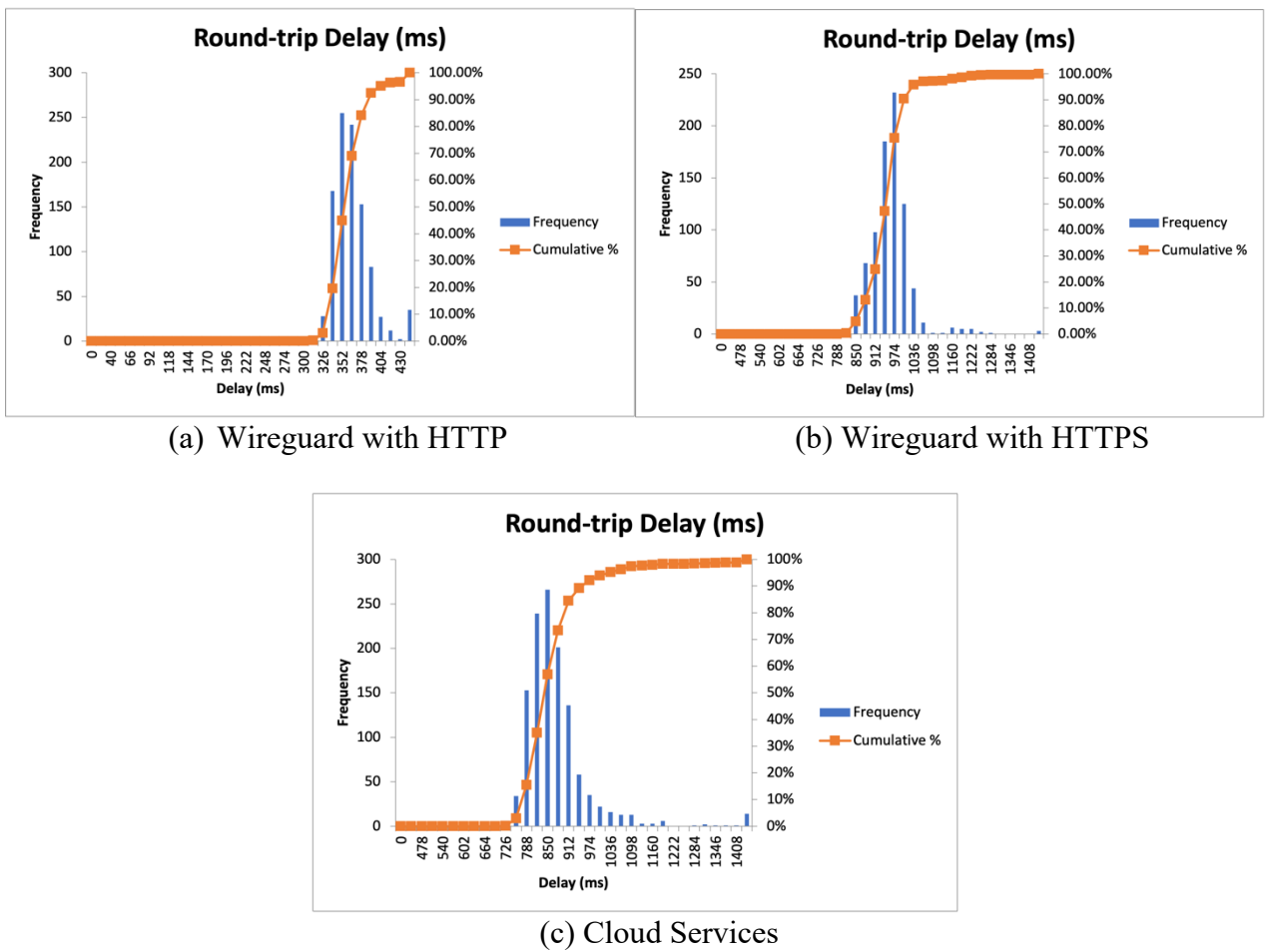

(a) Wireguard with HTTP

(b) Wireguard with HTTPS

(c) Cloud Services

Fig. 8: Round-trip delay histogram and cumulative probability for remotely controlled light system via 4G network link.

For the Cloud-based scenario, we can see that a minimum delay of about 723 ms and significantly higher standard deviation. We note that this minimum delay although lower compared to using WG with HTTPS, but the standards deviation when using WG is very small (of about 80) compared the measured one for this scenario (i.e. of about 795). We also note that samples in this scenario is more positively skewed (i.e. more higher delay samples or worse performance) compared to when using WG. Interestingly, we observed a much higher dispersion in this scenario compared to using WG with HTTPS, where the sample variance observed was 633324 compared to only 6460. The delay recorded in this scenario is much higher compared to the minimum delay of only 27 ms shown in Table V (i.e. locally controlled system using HTTP).

Figures 8a and 8b shows the histogram and CD of the delay measurements of samples over 4G network when using WireGuard for establishing a secure tunnel into the home network with HTTP and HTTPS respectively (described in section V-C). This tunnel ends directly at the IoT subnet to control the Philips Hue Light directly by using the RESTful API on the hub. The minimum delay achieved in the two scenarios is ≈310 ms and 788 ms. The distribution follows roughly a normal distribution with a mean value of 369 ms and 963 ms respectively, and a slightly positive skewness (i.e. we can see a number of observations on the right side of the mean with high delays) higher for HTTP then for HTTPS. The CD shows that 95 % of the delay samples are below 404 ms for HTTP and 1,036 ms for HTTPS, this is much higher compared to using local control (scenario V-A) which is expected. The minimum delay in these two scenarios is higher than the minimum delay observed in scenario V-A of 28 ms, this is mainly due to the 4G link delay, which additional delay due to communicating with HTTPS over the secure tunnel using WireGuard.

Figure 8c shows the histogram and Cumulative Distribution of the delay measurements of samples collected by controlling the light using Philips Cloud Services via a 4G network. The IoT device is still hosted in its own subnet as part of the home network which is connected to the internet using a fast speed link while the client is using 4G networks to send commands. This figure shows that the samples obtained are spread over compared to those shown in figure 8b with most of the tail being on the positive side of the mean value. This is also reflected in the higher variance (and higher standard deviation). This could contribute to a much varying user experience as the delays are not tight compared to those recorded when using WG.

Evaluation of Control via Office Network in Scenarios V-E and V-F

The following results were collected while using the scenarios described in sections V-E and V-F. Table VIII shows the results for using WG ()with HTTP and HTTPS) and Cloud-based respectively. The best performance (lowest delay) is ≈ 117 ms for the WG-HTTP scenario, while the next best performance is ≈ 362 ms for the Cloud-based scenario. The mean value of delay follows a similar pattern with WG-HTTP having the best delay of ≈ 159 ms with standard deviation of ≈82 ms. Lower standard deviation indicates that the variation of performance observed by the end-user around the mean is minimum compared to other methods. The best responsive method clearly is WG-HTTP which also offers a very tight standard deviation compared to the Cloud-based.

| Delay (ms) | WG-HTTP | WG-HTTPS | Cloud |
|---|---|---|---|
| Mean | 158.84 | 472.27 | 465.81 |
| Standard Error | 2.57 | 1.88 | 10.1 |
| Median | 150.27 | 462.75 | 432.5 |
| Standard Deviation | 81.69 | 63.5 | 322.11 |
| Simple Variance | 6672.97 | 4032.26 | 103756.77 |
| Kurtosis | 147.61 | 179.69 | 226.85 |
| Skewness | 11.96 | 11.84 | 14.81 |
| Range | 1121.89 | 1120.6 | 5128.89 |
| Minimum | 117.34 | 413.68 | 362.03 |
| Maximum | 1239.23 | 1534.27 | 5490.92 |
| Confidence Level (95%) | 5.05 | 3.69 | 19.81 |

TABLE VIII: Statistics for round-trip delay for controlled Philips Hue light system using (WG and Cloud Services) via office network.

Figures 9a and 9b show the histogram and CD of the delay measurements of samples over office network when using WireGuard for establishing a secure tunnel into the home network with HTTP and HTTPS respectively (described in section V-E). This tunnel ends directly at the IoT subnet to control the Philips Hue Light directly by using the RESTful API on the hub. The minimum delay achieved in the two scenarios is ≈117 ms and 414 ms. The distribution follows roughly a normal distribution with a mean value of 179 ms and 472 ms respectively, and a slightly similar skewness for HTTP then for HTTPS. The CD shows that 95 % of the delay samples are below 177 ms which is a much better performance using WG-HTTP, compared to a 509 ms when using WG-HTTPS. The average delay is much higher compared to using local control (scenario V-A), but much better compared to the 4G (scenario V-C). The minimum delay in these two scenarios is higher than the minimum delay observed in scenario V-A of 28 ms, this is mainly due to the office link delay.

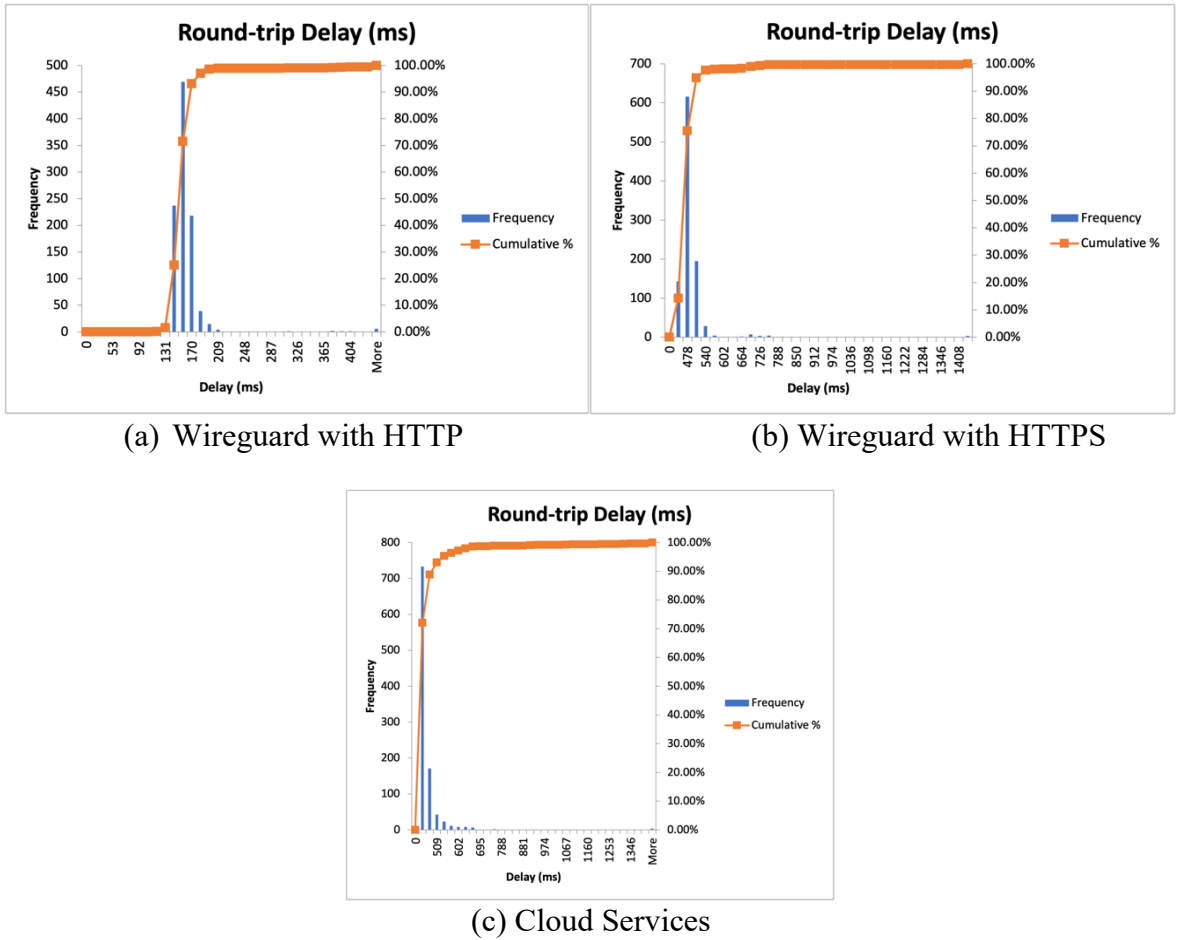

(a) Wireguard with HTTP  (b) Wireguard with HTTPS

(c) Cloud Services

Fig. 9: Round-trip delay histogram and cumulative probability for remotely controlled light system via office network.

HTTPS respectively (described in section V-E). This tunnel ends directly at the IoT subnet to control the Philips Hue Light directly by using the RESTful API on the hub. The minimum delay achieved in the two scenarios is ≈117 ms and 414 ms. The distribution follows roughly a normal distribution with a mean value of 179 ms and 472 ms respectively, and a slightly similar skewness for HTTP then for HTTPS. The CD shows that 95 % of the delay samples are below 177 ms which is a much better performance using WG-HTTP, compared to a 509 ms when using WG-HTTPS. The average delay is much higher compared to using local control (scenario V-A), but much better compared to the 4G (scenario V-C). The minimum delay in these two scenarios is higher than the minimum delay observed in scenario V-A of 28 ms, this is mainly due to the office link delay.

Figure 9c shows the histogram and Cumulative Distribution of the delay measurements of samples collected by controlling the light using Philips Cloud Services via office network (described in section V-F). It is notable that the minimum delay obtained was 362 ms which is better than when using WG-HTTPS (about 414 ms), which could be attributed to the way the traffic is routed towards the Philips cloud services in one of the global data-centre and also indicates that the added delay due to HTTPS encryption is one of the dominant factors. On the other hand, the figure shows that the cloud-based samples are more spread over compared to those shown in figure 9b with a higher positive skewness forming the tail on the positive side of the mean. This is also reflected in the higher variance of ≈103,757 (and higher standard deviation 322 ms compared to both WG-HTTP and WG-HTTPS), i.e. varied user experience compared to when using WG.

Evaluation of Control via Public WiFi in Scenarios V-G and V-H

The following results were collected while using the scenarios described in sections V-G and V-H. Table IX show the results for using WG ()with HTTP and HTTPS) and Cloud-based respectively. The best performance (lowest delay) is ≈ 115 ms for the WG-HTTP scenario, while the next best performance is ≈ 389 ms for the Cloud-based scenario. The mean value of delay follows a similar pattern with WG-HTTP having the best delay of ≈ 145 ms with standard deviation of ≈118 ms. Although the cloud-based scenario has a higher mean value of 477 ms, but it has a tighter standard deviation at only 94 ms. In this scenario also, the best responsive method is still WG-HTTP which also offers a tighter standard deviation compared to WG-HTTPS scenario and slightly higher than the Cloud-based scenario.

| **Delay (ms)** | **WG-HTTP** | **WG-HTTPS** | **Cloud** |
|---|---|---|---|
| **Mean** | 145.18 | 475.68 | 477.24 |
| **Standard Error** | 3.66 | 6.32 | 2.97 |
| **Median** | 136.96 | 464.48 | 455.26 |
| **Standard Deviation** | 118.23 | 200.69 | 94.46 |
| **Simple Variance** | 13977.88 | 40278.38 | 8922.14 |
| **Kurtosis** | 573.91 | 970.92 | 54.81 |
| **Skewness** | 22.28 | 30.88 | 6.39 |
| **Range** | 3195.42 | 6369.98 | 1167.12 |
| **Minimum** | 113.85 | 410.47 | 389.07 |
| **Maximum** | 3309.27 | 6780.45 | 1556.19 |
| **Confidence Level (95%)** | 7.18 | 12.41 | 5.83 |

TABLE IX: Statistics for round-trip delay for controlled Philips Hue light system using (WG and Cloud Services) via Public WiFi

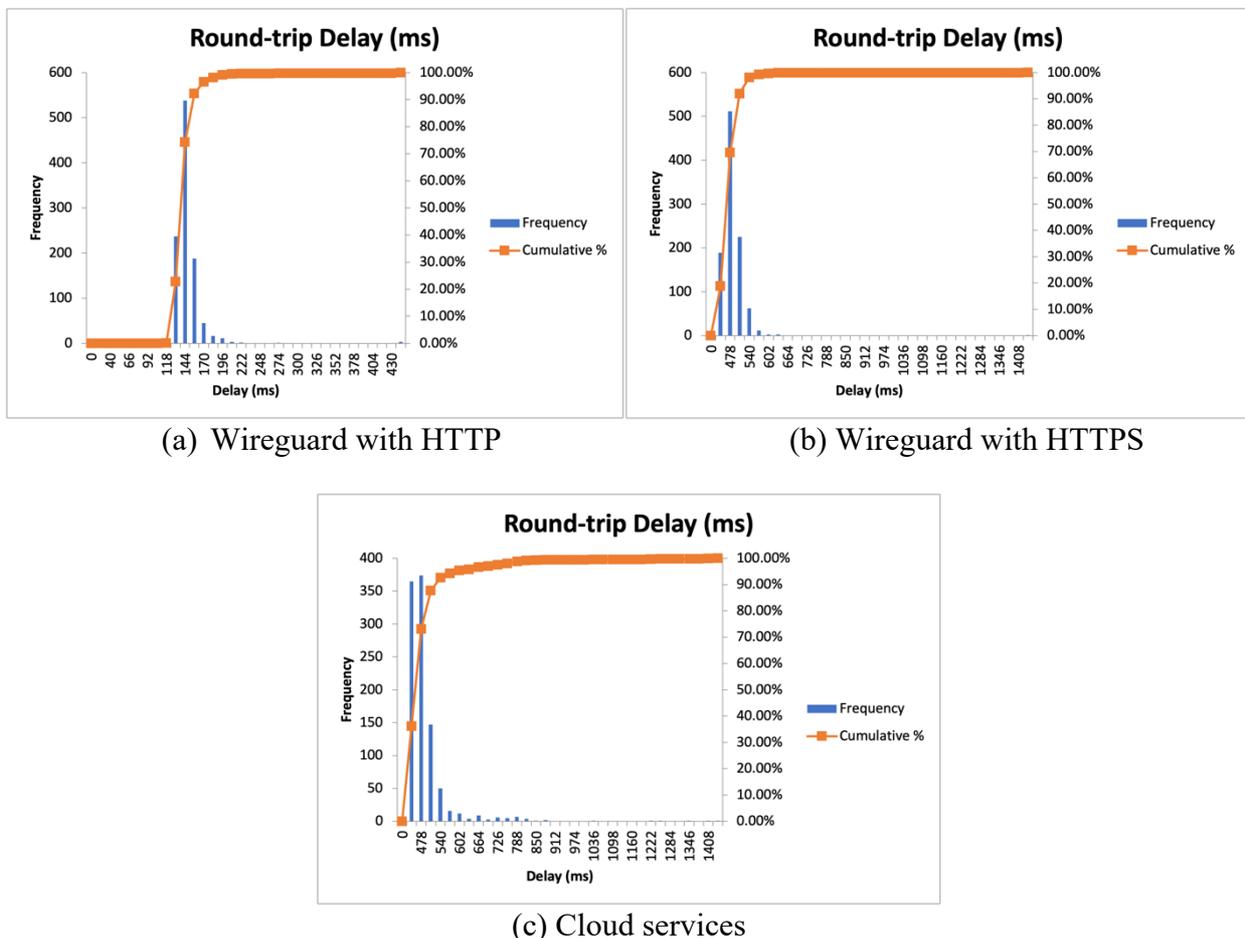

(a) Wireguard with HTTP

(b) Wireguard with HTTPS

(c) Cloud services

Fig. 10: Round-trip delay histogram and cumulative probability for remotely controlled light system via Public WiFi

Figures 10a and 10b show the histogram and CD of the delay measurements of samples over a public WiFi network when using WireGuard for establishing a secure tunnel into the home network with HTTP and HTTPS respectively (described in section V-G). Similar to previous scenarios, the tunnel ends directly at the IoT subnet to control the Philips Hue Light directly by using the RESTful API on the hub. The minimum delay achieved in the two scenarios is ≈114 ms and 410 ms. The distribution follows roughly a normal distribution with a mean value of 145 ms and 476 ms respectively, and a slightly higher positive skewness for HTTPS over HTTP. The CD shows that 95 % of the delay samples are below 164 ms for HTTP and 525 ms for HTTPS, which matches trends observed in previous scenarios. Both scenarios climb quickly and offer tighter distribution around the mean, compared to the cloud-based scenario. The minimum delay observed in these two scenarios is higher than the minimum delay observed in scenario V-A of 28 ms, but the public WiFi link offers delay that is consistent and low when using WG-HTTP.

Figure 10c shows the histogram and Cumulative Distribution of the delay measurements of samples collected by controlling the light using Philips Cloud Services via public WiFi network (described in section V-H). It is notable that the minimum delay obtained was 389 ms which is better than when using WG-HTTPS (about 410 ms) which indicates that the added delay due to HTTPS encryption is one of the dominant factors causing the delay. Contrary to some of the previous scenario, the cloud-based offers lower variance of ≈8,922 (and lower standard deviation 94 ms compared to both WG-HTTP and WG-HTTPS) which shows that results are tighter around the higher mean delay. This contributes to a consistent user experience of higher delays communicating through the Philips hue cloud services.

## VII.     CONCLUSION

With the wide deployment of IoT devices and the recent trend to moving from home, various security and privacy issues arise from allowing these devices to access or be accessed through the Internet. In this paper, we evaluate the effectiveness of existing enterprise solutions. We design an alternative domestic network hierarchy where IoT devices are isolated in a separate VLAN. We show a WireGuard VPN-based remote access solution to control IoT device from outside the home without requiring a TTP. We present a detailed evaluation the delays associated with this alternative design. In our work, we demonstrate that: 1) VPN-based direct communication between mobile device and IoT is achievable and offers both better privacy and lower latency compared to the cloud-based solution. 2) It is possible to completely remove the TTP as a middle-man from the loop (which reduces the information exposure to third parties as well security risks for many consumers by preventing vulnerable devices from being attacked by outsider adversary). 3) Isolating IoT devices in the domestic network is possible without degrading their performance (which also offers better control and prevents rogue IoT from compromising other areas of the network). As a future work, we aim to evaluate and measure performance for different types of consumers' IoT devices as well as deployment and evaluation within domestic settings to assess users' acceptance and feedback.

## ACKNOWLEDGEMENTS

This work was supported by the PETRAS 2: EP/S035362/1 National Centre of Excellence for IoT Systems Cybersecurity under the project "Tangible Security" (TanSec).